\documentclass{iopart}
\usepackage{epsf}
\usepackage{latexsym}
\begin{document}

\newcommand{\tbox}[1]{\mbox{\tiny #1}} 
\newcommand{\bbox}[1]{\mathbf{#1}}
\newcommand{\half}{\scriptstyle\frac{1}{2}\displaystyle}
\newcommand{\parl}{\mbox{\tiny $| \! |$}}


\title{Quantal Brownian Motion - \\ 
Dephasing and Dissipation}
\author{
\rm Doron Cohen \\ 
{\it Department of Physics of Complex Systems}, \\ 
{\it The Weizmann Institute of Science, Rehovot 76100, Israel.} \\
(Published in Journal of Physics A {\bf 31}, 8199-8220 (1998)))
}


\begin{abstract}
We analyze quantal Brownian motion in $d$ dimensions 
using the unified model for diffusion localization and 
dissipation, and Feynman-Vernon formalism. At high 
temperatures the propagator possess a Markovian property 
and we can write down an equivalent Master equation. 
Unlike the case of the Zwanzig-Caldeira-Leggett model,  
genuine quantum mechanical effects manifest themselves 
due to the disordered nature of the environment. 
Using Wigner picture of the dynamics we distinguish 
between two different mechanisms for destruction of 
coherence: scattering perturbative mechanism and 
smearing non-perturbative mechanism. 
The analysis of dephasing is extended to 
the low temperature regime by using a semiclassical 
strategy. Various results are derived for ballistic, 
chaotic, diffusive, both ergodic and non-ergodic motion. 
We also analyze loss of coherence at the limit of 
zero temperature and clarify the limitations of the 
semiclassical approach. The condition for having 
coherent effect due to scattering by low-frequency 
fluctuations is also pointed out. It is interesting that 
the dephasing rate can be either larger or smaller than 
the dissipation rate, depending on the physical circumstances.  
\end{abstract}

\section{Introduction}

Classical Brownian motion is 
described by the Langevin equation 
\begin{eqnarray} \label{e1}
m\ddot{\bbox{x}}+\eta\dot{\bbox{x}}={\cal F}
\end{eqnarray}
where $\bbox{x}$ is the position of the particle,  
$\eta$ is the friction coefficient, and    
${\cal F}=-\nabla{\cal U}(\bbox{x},t)$ is 
a stochastic force. This equation is 
meaningful only in a statistical sense. 
The time evolution of a phase space 
distribution $\rho(\bbox{x},\bbox{p})$ is obtained by solving 
(\ref{e1}) for various realizations of 
the stochastic potential, and then averaging 
over all these realizations. The  
stochastic potential is zero on the 
average and its correlations are 
\begin{eqnarray} \label{e2} 
\langle{\cal U}(\bbox{x}'',t''){\cal U}(\bbox{x}',t')\rangle
=\phi(t''{-}t')\cdot w(\bbox{x}''{-}\bbox{x}') 
\end{eqnarray}
Typically these correlations are 
characterized by a `short' temporal 
scale $\tau_c$ and a `microscopic' 
spatial scale $\ell$. 
Usually, it is further assumed that 
higher moments are determined 
by Gaussian statistics.  The Langevin 
description can be derived by considering 
a general Hamiltonian of the form 
\begin{eqnarray} \label{e3}
{\cal H} \ = \ {\cal H}_0(\bbox{x},\bbox{p})
+{\cal H}_{env}(\bbox{x},Q_{\alpha},P_{\alpha}) 
\end{eqnarray}
where $\bbox{x}$ and $\bbox{p}$ are the canonical variables 
that correspond to the distinguished degree 
of freedom, ${\cal H}_0=\bbox{p}^2/2m$ is the 
free motion Hamiltonian, and $(Q_{\alpha},P_{\alpha})$ are 
environmental degrees of freedom.
The dynamical variable $\bbox{x}$ may represent 
the position of a large particle. In the 
one-dimensional version of Brownian motion it 
may represent the position of a piston.   
The actual conditions for having the reduced 
Langevin description turn out to be quite 
weak~\cite{wilk}. The environment should consist of at least $3$ degrees 
of freedom with {\em fast} chaotic dynamics.  {\em Fast} 
implies that the classical motion is characterized by 
a continuous spectrum with high frequency cutoff, such 
that the motion of the environment can be treated 
adiabatically with respect to the {\em slow} motion of 
the particle. It is essential to assume 
that the following condition is fulfilled 
\begin{eqnarray} \label{e4}
\mbox{Generic Brownian Motion} 
\ \ \ \  \Leftrightarrow \ \ \ \  
\frac{v}{\ell} \ll \frac{1}{\tau_c} 
\end{eqnarray}
where $\tau_c$ and $\ell$ characterize 
the correlations of the stochastic 
potential which is experienced by the particle. 
Eq.(\ref{e4}) is the condition for using 
the white noise approximation (WNA). namely, 
the noise is characterized by its intensity 
\begin{eqnarray}   \label{e5}
\nu \equiv \int_{-\infty}^{\infty}\phi(\tau)d\tau \cdot |w''(0)|
\end{eqnarray}
and its temporal correlations 
are adequately described by 
the formal expression $\phi(\tau)=\nu\delta(\tau)$. 
Without loss of generality we assume 
from now on the normalization $w''(0)=-1$. 
The correlations of the stochastic force satisfy 
\begin{eqnarray}   \label{e6}
\langle{\cal F}(t){\cal F}(t')\rangle_{\mbox{at $x$}} \ = \ \phi(t{-}t') 
\end{eqnarray}
In the general case,~(\ref{e6}) is less informative than 
(\ref{e2}). However, in case of a classical particle that 
experiences white noise, the additional information 
is not required at all!  It should be 
emphasized that this observation does not 
hold if~(\ref{e4}) is not satisfied:  
If $\tau_c$ is larger than $\ell/v$, then  
the particle will perform a stochastic  
motion that depends crucially on the 
"topography" of the stochastic potential.

The classical analysis further reveals~\cite{wilk} 
that associated with the stochastic potential 
there is also a dissipation effect. If the 
environment is characterized by either  
micro-canonical or canonical temperature 
$T$, then the friction parameter will be 
\begin{eqnarray}   \label{e7}
\eta \ = \ \frac{\nu}{2k_BT}
\ \ \ \ \ \mbox{Fluctuations $\leadsto$ Dissipation} 
\end{eqnarray}
Thus, any generic environment, in the 
sense specified above, leads to the 
universal ohmic behavior. 
The motion of the particle is determined 
by the interplay between the friction and 
the noise. The friction leads to damping of the 
particle's velocity, while the noise pumps energy back 
into its motion. Eventually we have 
diffusion with coefficient $D_{\eta}=\nu/\eta^2$.   
If~(\ref{e4}) is not satisfied, then we 
will have diffusion even in the absence 
of dissipation.  The latter "non-dissipative 
diffusion" is characterized by the coefficient 
$D_0 \sim \ell v$, where $v$ is the velocity 
of the particle. This latter type of diffusive 
behavior should not be confused 
with the generic "dissipative diffusion" 
that characterizes Brownian motion.

One wonders whether there is a well defined 
quantized version of the above Langevin equation. 
Is it possible, quantum-mechanically, to characterize 
a universal dynamical behavior that corresponds to 
classical Brownian motion? Are stronger 
physical conditions required in order to 
guarantee generic behavior? 
In the present paper we have no intention to give 
a full answer to all these questions. 
Rather, as in previous publications~\cite{dld}, 
we follow the Caldeira-Leggett strategy~\cite{CL}. 
Namely, we consider the motion of a particle 
under the influence of an effective (non-chaotic) 
bath that is composed of infinitely many oscillators. 
The proper model that corresponds (classically) to 
the Langevin equation~(\ref{e1}) with~(\ref{e2}) 
is the DLD model that has been introduced in~\cite{dld}.
DLD are the initials of ``Diffusion Localization and Dissipation''. 
These three effects comes out naturally from the 
quantum-mechanical solution of the DLD model. 

Quantum mechanically, it is convenient to use 
Wigner function $\rho(R,P)$ in order to represent
the reduced probability density matrix. 
We can use the Feynman-Vernon (FV) formalism~\cite{FV} 
in order to find an expression for the 
propagator ${\cal K}(R,P|R_0,P_0)$. This propagator 
is uniquely determined once the friction ($\eta$), 
the temperature of the bath ($T$) and the spatial   
correlations $w(r)$ are specified. 
The dependence of the propagator on the bath-temperature 
is via the appearance of a quantum-mechanical version of 
the noise kernel $\phi(\tau)$.  
The latter is related to the friction coefficient  
via a universal fluctuation-dissipation relation. 
At high temperatures the quantum-mechanical 
$\phi(\tau)$ coincides with its classical expression. 
At the limit of zero temperature  $\phi(\tau)$ 
does not vanish, rather it develops a large negative tail.
In the latter case, $\phi(\omega)$, the power spectrum of 
the noise, reflects the zero point fluctuations (ZPF) 
of the environmental modes. 

The Zwanzig-Caldeira-Leggett (ZCL) model~\cite{CL} 
constitutes a special formal limit of the DLD model. 
It is obtained by taking the limit 
$\ell\rightarrow\infty$. Both the ZCL model 
and the DLD model will be generalized in this 
paper to allow the analysis of Brownian motion 
in $d$-dimensions.  A modified version of the 
the DLD model incorporates the effect of long 
range (power-law) spatial correlations.  
On the other hand, we are not discussing 
the DLD model in its full generality (as in~\cite{dld}),  
but rather restricting ourselves to the particular 
circumstances that correspond in the classical 
limit to generic Brownian motion. Thus, we 
are considering the {\em ohmic} DLD model, and 
further assume that~(\ref{e4}) is satisfied.

In case of the ZCL model, the quantal propagator 
${\cal K}(R,P|R_0,P_0)$ is a Gaussian 
stochastic kernel. Consequently the dynamics 
can be obtained by solving the Langevin 
equation~(\ref{e1}) with~(\ref{e6}).
At low-temperatures the quantum mechanical 
version of $\phi(\tau)$ should be used. 
At high temperatures $\phi(\tau)$ becomes 
classical-like and consequently the ZCL propagator  
coincides with its classical limit. 
These observations do not hold in case of the 
DLD model~\cite{dld}.  
Furthermore, the distinction between the 
quantal DLD propagator and its classical limit 
persists even in the limit of high temperatures.

It is important to define what is 
the meaning of `high temperatures'.  
As in the classical case, a relatively 
simple description of the dynamics is 
obtained if it possesses a Markovian property. 
With a Markovian property it is possible 
to obtain the long-time evolution by composing 
short-time evolution steps. It is 
also possible then to write down a corresponding 
Master equation for the Wigner function.  
The FV path-integral expression for the 
propagator contains the quantum-mechanical 
version of $\phi(\tau)$ rather than the 
classical one. 
In order to have a Markovian property one 
should argue that it is possible to 
use the WNA, meaning to replace 
$\phi(\tau)$ by the effective classical-like 
delta-function. 
Still, in case of the DLD model, the result 
of the path integration is not classical-like: 
The high-temperature Markovian limit is 
not the same as the classical limit.

It turns out that the quantum-mechanical 
condition for using the WNA, thus having 
a Markovian property, is more 
restrictive than~(\ref{e4}). For ballistic-like 
motion we shall see that the actual condition is,   
in most circumstances,   
\begin{eqnarray} \label{e8}
\mbox{High Temperatures} 
\ \ \ \  \Leftrightarrow \ \ \ \  
\frac{v}{\ell} \ll \frac{k_B T}{\hbar} 
\end{eqnarray}
A recent derivation of the high temperature 
ohmic DLD model which is based on a synthetic RMT 
Hamiltonian has been reported in~\cite{rmt}. 
The existence of such a derivation is most encouraging 
since it further support the idea of having 
a universal description of quantal Brownian 
motion, at least in the Markovian limit. 
It has been speculated by the authors 
of the latter reference that a future extension 
of their formalism will lead to an agreement 
with the general result of~\cite{dld}.  
The high temperature ohmic DLD model has been 
discussed in~\cite{dld} and has been 
further analyzed in~\cite{rmt}. The 
physical picture of high-temperature dynamics 
will be further illuminated in the present paper.

Loss of coherence, or dephasing, is a central 
issue in the analysis of quantal Brownian motion. 
Wigner's picture of the dynamics leads to the distinction 
between two different mechanisms for loss of coherence. 
In case of ZCL model one should consider 
the `{\em spreading mechanism}'. This classical-like 
non-perturbative mechanism is very 
effective in smearing away the interference pattern. 
In case of the DLD model, coherence is much better 
maintained, and one should consider the 
perturbative `{\em scattering mechanism}' for dephasing.

At the limit of zero temperature, the spreading 
mechanism is still effective in suppressing interference. 
One can use the Langevin formalism in order to 
analyze the smearing of the interference pattern. 
It is important, however, to take into account the negative 
temporal-correlations of the effective noise~\cite{noise}.  
On the other hand, the analysis of low-temperature dephasing 
in case of the `scattering mechanism' is a quite subtle issue,   
that constitutes a main concern of the present paper.  
The Langevin formalism is no-longer applicable, and the lack 
of a Markovian property enforce a semiclassical approach. 
The semiclassical approach has a further advantage: 
it is possible to go beyond the analysis of 
a simple ballistic-like Brownian motion and to 
analyze other types of transport. We shall 
distinguish between ballistic, diffusive and chaotic 
motions through cavities. We shall derive 
various results and contributions to the 
dephasing rate in the various temperature regimes 
and depending on the physical circumstances.     
Some of these results coincide with similar 
computations that are related to electrons in 
metal~\cite{imry}.

An important question is whether the scattering mechanism 
is still effective in suppressing interference at the 
limit of zero temperature \cite{webb,dph}. 
It turns out that our semiclassical approach, 
in spite of its other advantages, has a limited 
range of validity. It can be trusted if 
the kinetic energy of the particle is sufficiently 
large. For ballistic-like motion large energy means 
\begin{eqnarray} \label{e8b}
\mbox{Large Energy} 
\ \ \ \  \Leftrightarrow \ \ \ \  
\frac{v}{\ell} \ll \frac{E}{\hbar} 
\end{eqnarray}
in analogy with~(\ref{e8}). The latter condition can be 
cast into the more suggestive form $\lambda_B\ll\ell$, 
where $\lambda_B $ is the De-Broglie wavelength of the 
particle. The large-energy condition is obviously  
satisfied in case of the ZCL model ($\ell\rightarrow\infty$). 
The large energy condition may not be satisfied  
in case of the DLD model, and consequently the 
semiclassical result should be modified.  
In particular, in case of a low-temperature 
thermal motion, the contribution of the 
zero point fluctuations (ZPF) to the dephasing rate 
should be excluded. 


\newpage

\section{The ohmic DLD Model}

We consider a bath that consists of infinitely 
many oscillators whose canonical coordinates 
are $(Q_{\alpha},P_{\alpha})$. The bath 
Hamiltonian is  
\begin{eqnarray} \label{e9} 
{\cal H}_{bath} \ = \ \sum_{\alpha}\left
(\frac{P_{\alpha}^2}{2m_{\alpha}}
+\frac{1}{2} m \omega_{\alpha}^2 Q_{\alpha}^2\right)
\end{eqnarray}
In case of the DLD model the interaction of the 
particle with the oscillators is described by  
\begin{eqnarray} \label{e10}
{\cal H}_{int} = - \sum_{\alpha} c_{\alpha} Q_{\alpha} 
u(\bbox{x}{-}\bbox{x}_{\alpha})
\end{eqnarray}
where $\bbox{x}_{\alpha}$ is the location of the 
$\alpha$ oscillator, $u(\bbox{x}-\bbox{x}_{\alpha})$ 
describes the interaction between the particle 
and the $\alpha$ oscillator, and $c_{\alpha}$ 
are coupling constants. It is assumed that the 
function $u(\bbox{r})$ depends only on $|\bbox{r}|$. 
The range of the interaction will be denoted 
by $\ell$. 
The oscillators are distributed uniformly all 
over space. Locally, the distribution of their 
frequencies is ohmic.  Namely, 
\begin{eqnarray}    \label{e11}
\frac{\pi}{2} \sum_{\alpha}
\frac{c^2_{\alpha}}{m_{\alpha}\omega_{\alpha}} 
\delta(\omega-\omega_{\alpha}) \ 
\delta(\bbox{x}-\bbox{x}_{\alpha})
\ = \ \eta\omega  \ \ \ \ 
\mbox{for $\omega<1/\tau_c$} \ \ \ \ .
\end{eqnarray}
It is useful to define the 
spatial auto-correlation function
\begin{eqnarray} \label{e12}
w(\bbox{r})=\int_{-\infty}^{\infty}
u(\bbox{r}{-}\bbox{x}')u(\bbox{x}')dx'
\end{eqnarray}
Without loss of generality $u(\bbox{r})$ is 
normalized such that $w''(0)=-1$. For example,  
we may consider a Gaussian $u(\bbox{r})$ for 
which 
\begin{eqnarray}    \label{e13}
w(\bbox{r}) \ = \ \ell^2\exp
\left(-\frac{1}{2}\left(\frac{\bbox{r}}{\ell}\right)^2\right)  
\end{eqnarray}
The $d$-dimensional FT of $w(\bbox{r})$ will be 
denoted by $\tilde{w}(\bbox{k})$. The  
mode-density (after angular integration) 
is $g(k) = (C_d/(2\pi)^d) k^{d{-}1} \tilde{w}(k)$.
In general, we shall assume that
\begin{eqnarray} \label{e14}  
g(k)=C \ell^{2+\sigma} k^{\sigma{-}1}
\ \ \ \ \mbox{for} \ \ k<1/\ell
\end{eqnarray}
where $\ell$ characterize the spatial scale of the 
correlations, and $C$ is a dimensionless constant.
In case of the short range Gaussian-type 
correlations~(\ref{e13}), the parameter $\sigma$ equals 
simply to the dimensionality $d$. For  
long range power-law interaction it may be less than $d$, 
possibly negative. In a moment we shall argue that in   
order to have a well defined model we must have 
$|w''(0)|<\infty$. Therefore only $-2<\sigma$ is meaningful. 
The regime $-2<\sigma \le 0$ is well defined but it requires 
special treatment since $w(0)$ diverges.

The formal limit $\ell\rightarrow\infty$ 
corresponds to the ZCL model. The ZCL model 
describes interaction of a particle with 
environmental modes whose wavelength is 
much larger compared with the distances 
that are explored by the particle. 
Consequently, the interaction with 
the $\alpha$ field-mode is approximated by 
a linear function. The interaction term
in the Hamiltonian that {\em defines} the 
ZCL model is of the following form 
(here generalized to $d$-dimensions):
\begin{eqnarray} \label{e15}
{\cal H}_{int} \ = \ \sqrt{d} 
\sum_{\alpha} c_{\alpha}Q_{\alpha} 
\hat{n}_{\alpha} {\cdot} \bbox{x}
\end{eqnarray}
The unit vectors $\hat{n}_{\alpha}$
are assumed to be distributed isotropically.
The distribution of the oscillators with 
respect to their frequencies is assumed 
to be ohmic, as in~(\ref{e11}) with the 
$\delta(\bbox{x}{-}\bbox{x}_{\alpha})$ omitted.

The classical analysis of the $d$-dimensional 
DLD model constitutes a trivial generalization 
of the one-dimensional DLD model that has been 
considered in~\cite{dld}.  Equation (3.10) 
in the latter reference should be replaced 
by the following expression for the friction:  
\begin{eqnarray}  \label{e16}
{\cal F}_{friction}(\bbox{v}) \ = \
\int_{-\infty}^{\infty}\alpha(\tau) \ 
\nabla w(\bbox{v}\tau) \ d\tau  
\end{eqnarray}
where $\bbox{v}$ is the velocity of the particle. 
For ohmic bath $\alpha(\tau)$ can be replaced 
by $-\eta\delta'(\tau)$. One obtains 
then ${\cal F}_{friction}=-\eta\bbox{v}$.  
Obviously, in order to obtain a finite 
result it is crucial to have 
$|w''(0)|<\infty$. Recall that by convention  
we use the normalization $w''(0)=-1$.  
Thus, the generalized version~(\ref{e14}) of 
the DLD model is well-defined provided $-2<\sigma$.  
The reduced motion of the particle 
will be described by the Langevin 
equation~(\ref{e1}), where the noise 
satisfies~(\ref{e2}), with 
$\phi(\tau)=2\eta k_B T\delta(\tau)$. 
For the realization of a classical 
trajectory, the global functional-form 
of $w(\bbox{r})$ is of no importance.      
Only the force correlations~(\ref{e6}) are 
important.  The latter correlations  
are well defined as long as $|w''(0)|<\infty$, 
and by our convention they are equal $\phi(\tau)$. 
The parameter $\ell$ is insignificant in 
the classical analysis of the ohmic DLD 
model, and therefore the classical description 
of Brownian motion is the same as in the 
case of the ZCL model.      

Quantum mechanically, 
the reduced dynamics of the system may be described 
by the propagator ${\cal K}(R,P|R_0,P_0)$ of the probability 
density matrix. For sake of comparison with the classical limit 
one uses Wigner function $\rho(R,P)$ in order to represent the   
latter. Using Feynman-Vernon (FV) formalism~\cite{FV}, 
an exact path-integral expression for this propagator is 
obtained. The FV expression is a double sum 
$\int\int{\cal D}x'{\cal D}x''$ over the path variables 
$x'(\tau)$ and $x''(\tau)$. It is convenient to use 
new path variables $R=(x'{+}x'')/2$ and  $r=(x''{-}x')$, 
and to transform the $\int\int{\cal D}R{\cal D}r$ integral   
into the form~\cite{dld}
\begin{eqnarray} \label{e17}  
{\cal K}(R,P|R_0,P_0) \ \ = \ \ 
\int_{R_0,P_0}^{R,P} {\cal D}R \ \ {\cal K}[R]  \ \ \ \ ,
\end{eqnarray}
where ${\cal K}[R]$ is a real functional, which is defined 
by the expression: 
\begin{eqnarray}  \label{e18}  
{\cal K}[R] =
\int {\cal D}r 
\ \ \mbox{e}^{i\frac{1}{\hbar}(S_{free}+S_F)} 
\ \ \mbox{e}^{-\frac{1}{\hbar^2}S_N}
\ \ \ \ \ .
\end{eqnarray}
The ${\cal D}r$ integration is unrestricted at the endpoints,
and the free action functional is 
$S_{free}[\bbox{R},\bbox{r}]=-m\int_0^td\tau\ \ddot{R}r$. 
The action functional $S_F[\bbox{R},\bbox{r}]$ corresponds 
to the friction, and the functional $S_N[\bbox{R},\bbox{r}]$ 
corresponds to the noise. The explicit expression for the 
friction action functional, assuming an 
ohmic bath, is 
\begin{eqnarray}   \label{e19}
S_F[\bbox{R},\bbox{r}] \ = \ \eta \int_0^t d\tau \ 
\nabla w(\bbox{r}) \cdot \dot{\bbox{R}}
\end{eqnarray}
The general expression for the noise 
functional is 
\begin{eqnarray}  \nonumber
S_N[\bbox{x}',\bbox{x}'']=\frac{1}{2} \int_0^t\int_0^t dt_1 dt_2 
\ \phi(t_2{-}t_1) \ \times \ \ \\
\ \ \ \ [w(\bbox{x}_2''{-}\bbox{x}_1'')+
w(\bbox{x}_2'{-}\bbox{x}_1')-2w(\bbox{x}_2''{-}\bbox{x}_1')]
\label{e20} 
\end{eqnarray}
where $\bbox{x}_i$ is a short notation for $\bbox{x}(t_i)$.
The power spectrum of the noise $\phi(\omega)$ is  
the FT of the noise kernel $\phi(\tau)$. For ohmic bath 
\begin{eqnarray} \label{e21}  
\phi(\omega) \ = \ \eta\omega 
\ \hbar\coth\left(\frac{\hbar\omega}{2k_BT}\right)
\ \ \ \ \mbox{for} \ \ \omega<1/\tau_c 
\end{eqnarray}
The ZCL version for $S_F$ and $S_N$ is obtained by 
taking the limit $\ell\rightarrow\infty$ which is 
equivalent to the formal substitution 
$w(\bbox{r})=-\bbox{r}^2/2$.

The power spectrum~(\ref{e21}) of the 
quantum mechanical noise is the same as that  
of a classical white noise in the  restricted 
frequency regime $\omega < k_B T / \hbar$, 
where $\phi(\omega)=2\eta k_BT$.
If the temperature is high enough, such that 
$1/\tau_c < k_B T / \hbar$, then 
the power spectrum is essentially classical. 
At lower temperatures the power spectrum 
contains a non-trivial frequency zone 
$k_B T / \hbar < \omega < 1/\tau_c$, where  
one may use the approximation 
$\phi(\tau)=\hbar\eta\omega$. 
This component of the power spectrum 
does not vanish in the limit of zero temperature.  
It corresponds to the "zero point fluctuations" (ZPF) 
of the environmental modes. 
The temperature becomes effectively zero 
if $k_B T / \hbar < 1/t$. At intermediate 
temperatures, namely  $1/t < k_B T / \hbar < 1/\tau_c$, 
it is convenient to write the power spectrum as a sum 
$\phi(\omega)=\phi_0(\omega)+\phi_T(\omega)$, 
where $\phi_0(\omega)$ corresponds to the ZPF and 
$\phi_T(\omega)$ is the excess thermal noise:
\begin{eqnarray}    \label{e42}
\phi_T(\omega) \ = \ 
2\eta\frac{\hbar|\omega|}
{\exp\left(\frac{\hbar|\omega|}{k_B T}\right)-1}
\ \ \ \ \mbox{for} \ \ \omega<1/\tau_c 
\end{eqnarray}
The propagator~(\ref{e17}) possess a Markovian 
property if it is legitimate to make the WNA.  
The actual condition for having effectively 
white noise will be discussed in a later section.
At low temperatures it is essential to consider  
the non-trivial nature the quantal noise   
in order to obtain the proper quantum-mechanical 
dynamical behavior~\cite{CF}.

\section{Propagation in High Temperature Environment}   
  
In the absence of noise and friction, the 
free-motion propagator of Wigner function is 
the same as in the classical limit. Namely, 
\begin{eqnarray} \label{e22}
{\cal K}(R,P|R_0,P_0) \ = \ 
{\cal K}_{\mbox{\tiny free}}^{(cl)} 
\ = \ 2\pi\delta(P{-}P_0) \ 
\delta((R{-}R_0)-\mbox{$\frac{P}{m}$}t)
\end{eqnarray}
For simplicity, but without loss of 
generality, we shall consider whenever 
possible one-dimensional motion. 
In case of the ZCL model, both 
$S_F$ and $S_N$ are quadratic forms, 
and therefore the propagator is 
a Gaussian kernel. It follows that 
the dynamics of Wigner function is 
purely stochastic, and it can be 
described by Langevin equation 
(\ref{e1}) with~(\ref{e2}) and~(\ref{e21}). 
In this sense, there are no genuine quantum 
mechanical effects that are associated with 
Brownian motion, as long as the ZCL model is used 
for its description. The situation is 
quite different in case of the DLD model. 
Here the propagator is non-Gaussian, 
and, in general, it may have non-stochastic 
features.

\begin{figure}[t] 
\begin{center}
\leavevmode 
\epsfysize=3.4in
\epsffile{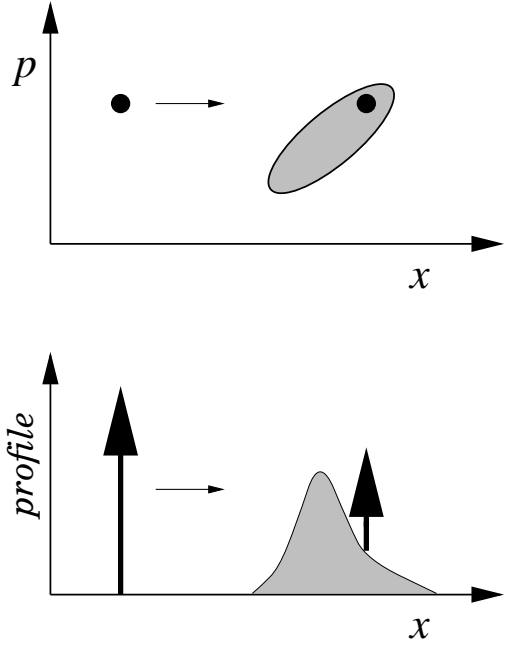}
\end{center}
\caption{\protect\footnotesize 
Upper plot: Phase space illustration for 
the structure of the DLD propagator. 
Lower plot: The projected phase-space density. } 
\label{f_propg}
\end{figure}

In the present section 
we shall restrict ourselves to the 
high temperature regime, where the 
white noise approximation (WNA), namely 
$\phi(\tau) = \nu \delta(\tau)$ can be 
applied. 
Whenever the WNA applies, the ZCL propagator 
becomes identical with the classical propagator 
for damped motion, namely
\begin{eqnarray} \label{e23}
{\cal K}(R,P|R_0,P_0) \ = \ 
{\cal K}_{\mbox{\tiny damped}}^{(cl)} 
\ = \ \mbox{Gaussian} 
\end{eqnarray}
In contrast, in case of the DLD model, the high 
temperature limit of the propagator 
{\em does not coincide} with its classical 
limit. An explicit calculation, as in~\cite{dld}, 
gives the following expression:   
\begin{eqnarray} \label{e24}
{\cal K} \ = \ 
W_{\hbar/\ell}\star {\cal K}_{\mbox{\tiny damped}}^{(cl)}
\ + \ \mbox{e}^{-\frac{2\eta k_BT\ell^2}{\hbar^2}t} 
(1{-}W_{\hbar/\ell}\star) \ {\cal K}_{\mbox{\tiny free}}^{(cl)}
\end{eqnarray}
This expression is valid also for the generalized 
DLD model provided $0<\sigma$. The precise value 
of the parameter $\ell$ is defined here via $w(0) \equiv \ell^2$.  
Above, $W(R-R',P-P')$ is a smooth Gaussian-like kernel that  
has unit-normalization. Its spread in phase space is 
characterized by the momentum scale $\hbar/\ell$, and 
by an associated spatial scale. The symbol $\star$ stands for 
convolution. Thus, the classical propagator is 
smeared on a phase-space scale that correspond to 
$\Delta p = \hbar/\ell$ and there is an additional 
un-scattered component that decay exponentially and 
eventually disappears. The structure of the propagator 
is illustrated in Fig.\ref{f_propg}. The significance 
of this structure will be discussed in the next section.  
  
To write down an explicit expression for the 
propagator is not very useful. Rather, it is 
more illuminating to follow the    
standard procedure~\cite{CL} and to write 
an equivalent Master equation. This is 
possible since at high temperatures the 
path-integral expression possess a Markovian property. 
Namely, since both $S_F$ and $S_N$ are local 
functionals of the paths, it is possible to 
regard the finite-time propagation as the 
convolution of (infinitesimal) short-time kernels. 
The derivation of the Master equation is 
explained in Appendix B. The final result is  
\begin{eqnarray} \label{e25}
\frac{\partial\rho}{\partial t} \ = \ \left[
\ - \ \partial_R \ \frac{1}{m} P
\ + \ \partial_P \ \frac{\eta}{m} G_F \star P
\ + \ \nu G_N \star  
\right] \ \rho
\end{eqnarray}
The friction kernel is defined as follows
\begin{eqnarray} \label{e26}
G_F \ \equiv \ {\cal FT} \left(\frac{w'(r)}{r}\right)
\ = \ \frac{1}{\hbar/\ell}
\hat{G}_F\left(\frac{P{-}P'}{\hbar/\ell}\right)
\end{eqnarray}
and the noise kernel is 
\begin{eqnarray} \nonumber
G_N \ \equiv \ \frac{1}{\hbar^2}
{\cal FT} \left( w(r){-}w(0) \right) 
\ = \\  \ \ \ \ \ \ \ \ \ \ 
\left(\frac{\ell}{\hbar}\right)^2
\left[ \frac{1}{\hbar/\ell}
\hat{G}_N\left(\frac{P{-}P'}{\hbar/\ell}\right) 
- \delta(P{-}P')\right] 
\label{e28.1}
\end{eqnarray}
Both $\hat{G}_F$ and $\hat{G}_N$ are 
smooth Gaussian-like scaling functions, 
properly normalized to unity.

If Wigner function does not possess fine 
details on the momentum scale $\hbar/\ell$, 
then the convolution with $G_F$ can 
be replaced by multiplication with $1$, 
and the convolution with $G_N$ can 
be replaced by ${\partial^2}/{\partial P^2}$. 
These replacements are formally legitimate 
both in the classical limit 
$\hbar \rightarrow 0$, and in the ZCL limit 
$\ell \rightarrow \infty$. One obtains 
then the classical Fokker Planck equation   
\begin{eqnarray} \label{e27} 
\frac{\partial\rho}{\partial t} \ = \ \left[
\ - \ \partial_R \ \frac{1}{m} P
\ + \ \partial_P \ \frac{\eta}{m} P
\ + \ \nu \frac{\partial^2}{\partial P^2}  
\right] \ \rho
\end{eqnarray}
The same observation can be done 
by inspection of~(\ref{e22}). Namely, 
if the propagator acts on a 
smooth Wigner-function, (no features 
on momentum scale  $\hbar/\ell$),  
then the second term becomes 
vanishingly small, while the first term 
becomes effectively classical.

\section{Dephasing Within the Wigner Picture}

\begin{figure}[t] 
\begin{center}
\leavevmode 
\epsfysize=3.4in
\epsffile{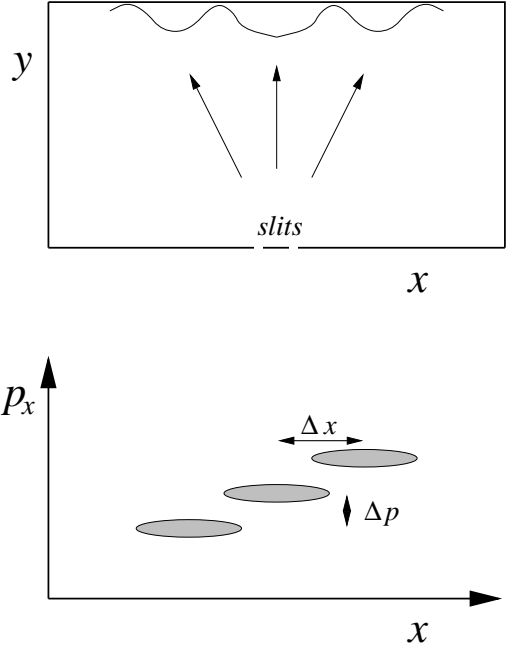}
\end{center}
\caption{\protect\footnotesize 
Upper plot: The geometry of a two-slit 
experiment. The propagation of the 
wavepacket is in the $y$ direction, and 
the interference pattern is resolved 
on the screen.  
Lower plot: Phase-space picture of the dynamics. 
Wigner function of the emerging wavepacket is projected 
onto the $(x,p_x)$ plane.} 
\label{f_slits}
\end{figure}

Wigner function may have some modulation on 
a fine scale due to an interference effect. 
The standard text-book example of a two slit 
experiment will be analyzed below, where 
the interference pattern has the momentum 
scale $\hbar/\bbox{d}$. See Fig.\ref{f_slits}.
$\bbox{d}$ is the distance between the slits. 
In this section we shall explain the possible 
mechanisms that lead to the disappearance of 
such interference pattern. 
In view of the semiclassical point of 
view of the next section, we use the term 
``dephasing'' in order to refer to this
disappearance effect.  
In the present section, as in the 
previous one, we are still limiting  
ourselves to the high temperature regime. 

In case of the ZCL model, the propagator 
is the same as the classical one, and therefore 
we may adopt a simple Langevin picture in 
order to analyze the dephasing process.      
Alternatively, we may regard the dephasing 
process as arising from Gaussian smearing of 
the interference pattern by the propagator.   
In case of the DLD model we should 
distinguish between two possible mechanisms 
for dephasing: 
\begin{itemize}
\item Scattering (Perturbative) Mechanism. 
\item Spreading (Non-Perturbative) Mechanism.
\end{itemize}
Actually, it is better to regard  
them as mechanisms to maintain coherence. 
The first mechanism to maintain coherence 
is simply not to be scattered  
by the environment. The second mechanism 
to maintain coherence is not to be 
smeared by the propagator. The first  
mechanism is absent in case of the   
ZCL model. We shall see shortly that the 
first mechanism is much more effective in 
maintaining coherence. The notions 
perturbative and non-perturbative are 
used in order to suggest a connection 
with an earlier work~\cite{noise}. The smearing 
effect can be regarded as arising from 
accumulation of many small-momentum 
scattering events. A conventional 
perturbative approach cannot be applied 
in order to analyze such dephasing process, 
even in the limit of very faint noise.

Having in mind the specific example of a 
two-slit experiment, we should distinguish 
between two cases.   
If the environment is characterized by 
a spatial correlation scale $\ell$ that 
is much larger than $\bbox{d}$, then we 
can `forget' about the DLD model and just 
use the ZCL model. The relevant mechanism 
for destruction of coherence in this case 
is the spreading mechanism.  
See detailed analysis in the next paragraph.
If the environment characterized by 
a spatial correlation scale $\ell\ll\bbox{d}$
then a totally different picture emerges. 
Now, Wigner function contains  
a modulation on a momentum scale 
much finer than $\hbar/\ell$. 
This modulation is not affected by the 
friction, but its intensity decays 
exponentially in time. This is based 
on inspection of either the propagator~(\ref{e22}), 
or the equivalent Master equation~(\ref{e25}). 
In the latter case note that 
the convolution with $G_N$ can be replaced 
by multiplication with $-(\ell/\hbar)^2$. 
The decay rate is     
\begin{eqnarray} \label{e28}
\frac{1}{\tau_{\varphi}}  
\ = \ \frac{2\eta k_BT\ell^2}{\hbar^2} 
\ \ \ \ \ \mbox{WNA for $0<\sigma$} 
\end{eqnarray}
It is a universal result for the dephasing 
rate due to the `scattering mechanism', 
since it does not depend on details of the 
quantum-mechanical state involved. 
However, the validity of this result is  
restricted to the high temperature 
regime, where the white noise approximation (WNA) 
can be applied.

For completeness we turn back to the detailed 
analysis of dephasing due to spreading. 
This mechanism leads to non-universal results, 
since the calculation of the dephasing 
time depends on actual details of the 
interference pattern.      
For concreteness we consider the simplest 
case of a two-slit interference experiment. 
The wavepacket that emerges from the 
two slits is assumed to be a superposition 
$\psi(\bbox{x})=\sum_{\pm}\varphi(|\bbox{x}|)
\exp(ik|\bbox{x}\mp\bbox{d}/2|)$, where 
$\hbar k$ is the momentum of the incident 
particle, and $\varphi(|\bbox{x}|)$ is 
a radial envelope function. The Corresponding 
Wigner function is  
\begin{eqnarray} \label{e29}
\rho(\bbox{x},\bbox{p}) \ \approx \ 
(1+\cos(\vec{\bbox{d}}{\cdot}\bbox{p}/4)) 
\times \rho_{\mbox{\tiny 1-slit}}(\bbox{x},\bbox{p}) \ = \
\sum_{n=-\infty}^{\infty} \rho_{n}(\bbox{x},\bbox{p})
\end{eqnarray}
The 1-slit Wigner function is multiplied 
by $\cos^2(\half \bbox{d} {\cdot} p_x)$,  
and therefore it is natural to regard 
the 2-slits Wigner function as composed 
of partial-wavepackets, each characterized   
by a different transverse momentum. 
By definition the partial-wavepacket $\rho_{n}$  
equals $\rho$ for $|p_x-2\pi n/\bbox{d}|<\pi/\bbox{d}$ 
and equals zero otherwise. Each partial wavepacket 
represents the possibility that the particle, being 
scattered by the slits,  
had acquired a transverse momentum which is 
an integer multiplet of $\Delta p_x = 2\pi\hbar/\bbox{d}$.  
Note that the corresponding angular separation 
is $\Delta p_x/(\hbar k)=\lambda_B/\bbox{d}$, as expected. 
the associated spatial separation is 
$\Delta x = (\Delta p /m){\cdot}t$ where 
$t=y/(\hbar k/m)$ is the time up to the screen. 
It is important to distinguish between the 
``preparation'' zone $y<\bbox{d}$ which is excluded 
from our considerations, and the far-field 
(Franhoufer) zone $\bbox{d}^2/\lambda_B \ll y$
where $\hbar \ll \Delta x\Delta p_x$  and 
consequently the individual partial-wavepackets 
can be resolved. Due to the noise the 
interference pattern is smeared on a momentum 
scale $\delta p_x = \sqrt{\nu t}$, and on 
a corresponding spatial scale 
$\delta x =  \sqrt{\nu t}{\cdot}t$. 
The interference pattern disappears once 
$\delta p_x \sim \Delta p_x$ or equivalently 
$\delta x \sim \Delta x$. This leads to 
the following expression for the dephasing time: 
\begin{eqnarray} \label{e30}
\frac{1}{\tau_{\varphi}}  
\ = \ \frac{\eta k_BT}{\hbar^2}d^2 
\ \ \ \ \ \mbox{WNA for ZCL model, two-slits} 
\end{eqnarray}
Comparing with~(\ref{e28}) we observe that indeed 
the smearing mechanism is very effective in 
suppressing the interference pattern. 
The result is also non-universal since it depends 
on details of the interference pattern. 
In some other circumstances, where Wigner 
function is characterized by a different 
type of interference pattern (notably the 
case of ``vertical'' interference pattern),  
the dephasing time may be proportional to 
some fractional power of the noise 
intensity~\cite{noise}. In general, in the 
full analysis of the smearing process, one should 
take into account the effect of friction. 
In the above example this effect has been 
neglected. 

The analysis of dephasing in case of the ZCL model 
is easily extended to the low-temperature regime.  
Langevin formalism is still applicable, provided 
the appropriate $\phi(\tau)$ is being used.  
The smearing of the interference pattern can be 
analyzed as in the previous paragraph. At the limit 
of zero temperature one should take into account 
the negative correlations of the noise~\cite{noise}. 
For the ballistic-like motion that has been considered 
above, the smearing scale $\delta p^2$ is proportional, 
at zero temperature, to $\ln(t)$ rather than to $t$, leading to 
an anomalous (non-exponential) dephasing factor~\cite{noise,dld}

\section{Dephasing - The Semiclassical Point of View}

\begin{figure}[t] 
\begin{center}
\leavevmode 
\epsfysize=1.4in
\epsffile{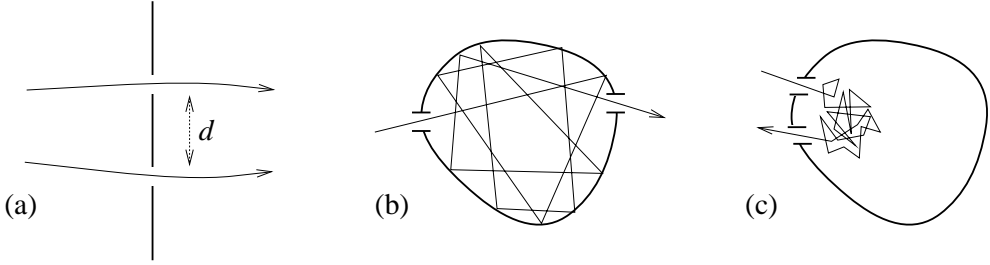}
\end{center}
\caption{\protect\footnotesize 
Various types of transport problems: (a) Ballistic 
transport as in the two-slit experiment; (b) Transport 
via a chaotic cavity;  (c) Transport via diffusive 
cavity, as in weak localization experiments. 
One should consider also the case of 
ergodic motion via a diffusive cavity (not plotted). } 
\label{f_trajc}
\end{figure}

There is a totally different approach 
to analyze dephasing, that is based on 
the semiclassical point of view. 
The advantage of the semiclassical approach is 
the ability to pursue the study of dephasing 
to the low-temperature regime as well as to 
circumstance in which the simple 
ballistic-like Brownian motion scenario 
is not applicable. Specific examples that 
will be discussed later are the 
transmission via either chaotic or diffusive 
cavity. From now on we regard the two-slit 
experiment as a special case of transport problem. 
See Fig.\ref{f_trajc}.  The probability 
to cross the obstacle, or more generally 
to be transmitted via some cavity, can be 
written as a (double) sum over classical 
trajectories. Namely,  
\begin{eqnarray} \label{e31}
\sum_{ab} A_a A_b^{*} 
\exp\left(
i\frac{S[\bbox{x}_a]{-}S[\bbox{x}_b]}{\hbar}
+i\frac{S_F[\bbox{x}_a,\bbox{x}_b]}{\hbar}
-\frac{S_N[\bbox{x}_a,\bbox{x}_b]}{\hbar^2}
\right) 
\end{eqnarray}
where $S[\bbox{x}_a]$ is the classical action 
for the classical trajectory $\bbox{x}_a$, 
and $A_a$ is the corresponding classical 
amplitude. Each pair $a\&b$ of classical trajectories 
constitutes a stationary-phase point of the exact 
path-integral expression. However, the notation is   
somewhat misleading since once the friction 
functional is switched on, $a$ and $b$ 
no longer can be considered independent indexes.  
In particular, strictly speaking, the 
diagonal terms are no longer truly diagonal. 
However, here comes a very important observation. 
In case of the DLD model, assuming that 
$\bbox{x}_a$ and $\bbox{x}_b$ are well separated 
with respect to the microscopic scale $\ell$, 
one has $S_F[\bbox{x}_a,\bbox{x}_b]=0$. 
Thus we conclude that friction has no effect 
on the interference part of the sum~(\ref{e31}), 
in consistency with our discussion 
of the scattering-mechanism in the 
previous section. On the other hand, friction
will have an effect on the diagonal terms of the 
sum, as required by the correspondence principle. 
It should be emphasized that in case of 
the ZCL model, friction may have a 
non-negligible effect also on the interference 
terms, again in consistency with our 
discussion of the smearing mechanism.

The suppression of an off-diagonal term in~(\ref{e31}) is given by 
the dephasing factor $\exp(-S_N[\bbox{x}_a,\bbox{x}_b]/\hbar^2)$. 
Using the definitions of Appendix D it is possible 
to transform expression~(\ref{e20}) into an integral 
over the Fourier components of the motion involved.     
Consequently, in most cases of interest, the dephasing factor can 
be cast into the form $\exp(-t/\tau_{\varphi})$, where  
\begin{eqnarray} \label{e37}  
\frac{1}{\tau_{\varphi}} = 
\frac{1}{\hbar^2} \int_0^{\infty} g(k)dk
\int_0^{\infty} \frac{d\omega}{\pi}  \phi(\omega)  P(k,\omega)
\end{eqnarray}
The domain of integration is illustrated in 
Fig.\ref{f_kwplan}.  The power spectrum 
$P(k,\omega)$ of the motion is calculated 
in Appendix D for ballistic, chaotic and 
diffusive motion. Various results that can 
be derived by using the above expression 
will be presented in later sections.

\begin{figure}[t] 
\begin{center}
\leavevmode 
\epsfysize=1.3in
\epsffile{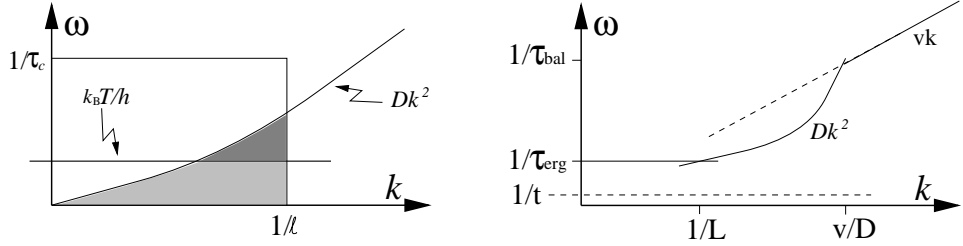}
\end{center}
\caption{\protect\footnotesize 
The $(k,\omega)$ plane. Left plot: The shaded 
regions indicate those environmental modes 
that are effective in the dephasing 
process. The darker region indicates a possible excess 
contribution due to ZPF. The curve $Dk^2$  
illustrates the frequency-span of $P(k,\omega)$. \   
Right plot: The power-spectrum of an ergodic motion 
via a diffusive cavity is concentrated under 
the plotted curves. See further details in Appendix D. } 
\label{f_kwplan}
\end{figure}

One should be very careful in the physical 
interpretation of $\exp(-t/\tau_{\varphi})$. 
If we have two unrelated trajectories $a\&b$, 
and we have also static disorder, 
then we will have "statistical" dephasing 
that reflects the effect of averaging over 
realizations of the disorder.   
Consider for example a two-slit experiment:  
The interference pattern on the screen will 
be distorted for any particular realization 
of a static disorder, and will be washed away
completely upon averaging over different realizations.   
This mechanism is non-effective if $a\&b$ are related 
by time reversal, for which static disorder gives 
$S_N[\bbox{x}_a,\bbox{x}_b]=0$. 
Still some suppression is expected
also in the latter case, via 
the classical-like amplitudes $|A_a|^2$.    
In case of dynamical environment the role of elastic 
scattering is taken by inelastic scattering events. 
Therefore we have genuine dephasing,  
irrespective of whether $a\&b$ are 
related by time reversal. In some 
typical circumstances, the dephasing 
factor can be re-interpreted as giving 
the probability to "leave a trace" in the 
environment (see Appendix C) . We take now the limit of zero 
temperature. Assuming for simplicity that 
$a\&b$ are related by time reversal, 
it follows from the definition of the influence 
functional that $\exp(-S_N[\bbox{x}_a,\bbox{x}_b]/\hbar^2)$ 
can be interpreted as the probability to 
excite at least one of the oscillators along the way. 

Up to now we have encountered, depending on the 
physical circumstances, two possible interpretations 
for the dephasing factor. The "statistical" interpretations 
holds in case of static disorder, while the 
"leaving-trace" interpretation holds in typical 
cases of a dynamical environment.    
Let us keep $T=0$ and take the limit $E\rightarrow 0$, 
where $E$ is the available energy of the particle.   
In order to make the following argumentation 
more illuminating, one may assume that the particle 
has a small but finite kinetic energy $E$, and that 
all the oscillators have relatively large frequencies, 
such that $E\ll\omega_{\alpha}$.   
Consequently, all the scattering events are elastic. 
On the other hand $1/\tau_{\varphi}$ has a non-vanishing value, 
which implies that the off-diagonals terms 
are being suppressed.  Obviously, the "leaving-trace" 
interpretation of the dephasing factor no longer holds.  
One wonders whether $\exp(-S_N/\hbar^2)$  acquires, 
under such circumstances, a somewhat different 
physical meaning, as in the case of static disorder. 
Maybe it gives now the probability to be scattered 
elastically (rather than inelastically) along the way, 
thus leading to dephasing of the "statistical" type.  
A perturbative treatment of the scattering process leads 
to the conclusion that the probability to be elastically scattered 
by a zero temperature oscillator is proportional to $c_{\alpha}^4$. 
This is because the elastic scattering off an oscillator 
involves a second-order process of virtual emission followed 
by absorption of a quanta $\hbar\omega_{\alpha}$. 
At the same time $S_N[\bbox{x}_a,\bbox{x}_b]$ is 
proportional to $c_{\alpha}^2$ to leading order. 
Therefore,  $\exp(-S_N/\hbar^2)$ cannot have the desired 
physical significance. 
We therefore must conclude that something goes 
wrong with the present semiclassical approach 
once low energy particle is concerned. We shall 
come back to this point later on.

\section{Dephasing at High Temperatures}

As a first step in the application of the 
semiclassical approach, it is interesting to recover 
results~(\ref{e28}) and~(\ref{e30}) for the dephasing 
rate. These results hold at high temperatures, 
such that the WNA is applicable. 
The noise functional in such circumstances 
takes the following simple form
\begin{eqnarray}    \label{e32}
S_N[\bbox{r}] \ = \  
2\eta k_BT \int_0^t \ [w(0){-}w(\bbox{r}(t'))] \ dt' 
\end{eqnarray}
where $\bbox{r}=\bbox{x}_a{-}\bbox{x}_b$. If the 
two trajectories are well separated with respect 
to the microscopic scale $\ell$, then one indeed recovers 
the universal result~(\ref{e28}). Recall that 
$w(0)\equiv\ell^2$. In the other extreme 
limit $\ell\rightarrow\infty$, that corresponds to 
the ZCL model, one obtains 
\begin{eqnarray}    \label{e33}
S_N[\bbox{r}] \ = \ 
2\eta k_BT \int_0^t \ \bbox{r}(t')^2 \ dt' 
\ \equiv \ 2\eta k_BT \ r_{\perp}^2 \cdot t
\end{eqnarray}
In case of the two-slit experiment $\bbox{r}\sim\bbox{d}$, 
and one recovers~(\ref{e30}). Obviously, it is not 
a universal result. For example, if the separation between 
the two trajectories grows linearly, then the dephasing time 
will be proportional to $1/T^{1/3}$ rather than to $1/T$.

The generalized DLD model with $-2<\sigma<0$, requires 
special care since $w(0)$ diverges. We are still  
considering high temperatures, such that the WNA is 
applicable. Using the notations of Appendix D  
expression~(\ref{e32}) can be cast into the form 
\begin{eqnarray}    \label{e34}
S_N[\bbox{r}] \ = \ 
\nu t \int w(\bbox{r}) P(\bbox{r}) d\bbox{r} 
\ = \ 
\nu t \int_0^{\infty} g(k)dk \ P(k)   
\end{eqnarray}
The latter expression converges for any 
$-2<\sigma$. In the ZCL limit, where 
$\ell$ is much larger than the average 
transverse distance between the trajectories, 
one may use the small-$k$ approximation of 
Appendix D, and then~(\ref{e33}) is recovered.   
In the DLD case, where $\ell$ is a very small scale,  
the integration should be split into the domains 
$k<k_{\perp}$ and $k>k_{\perp}$. The 
wavenumber $k_{\perp}\equiv 1/r_{\perp}$ is associate 
with the transverse distance $r_{\perp}$ between the  
two trajectories. Namely,   
\begin{eqnarray} \label{e35}
r_{\perp} \ \approx \ \left\{ \matrix{ 
\bbox{d}  & \mbox{for ballistic motion}  
            & t<\tau_{\tbox{bal}} \cr
\sqrt{Dt} & \mbox{for diffusive motion} 
            & \tau_{\tbox{bal}}<t<\tau_{\tbox{erg}} \cr
L         & \mbox{for ergodic motion} 
            & \tau_{\tbox{erg}}<t  
}\right.
\end{eqnarray}
Note that ergodic motion refers to either chaotic 
or diffusive trajectories with $\tau_{\tbox{erg}}<t$.  
Again we use the notations of Appendix D.  
For $0<\sigma$ the integration 
is dominated by the upper cutoff $k \sim 1/\ell$. 
One recovers then the standard result~(\ref{e28}). 
On the other hand, for $-2<\sigma<0$, the 
integration is dominated by 
the $k \sim k_{\perp}$ modes. The wavelength 
of these modes is like the transverse distance 
between the trajectories. One obtains then
\begin{eqnarray}    \label{e36}
S_N[\bbox{r}] \ = \ 
2\eta k_B T \ \ell^2  
\left( \frac{\ell}{r_{\perp}}\right)^{\sigma} \cdot t
\ \ \ \ \ \mbox{WNA for $-2<\sigma<0$}
\end{eqnarray}
For non-ergodic diffusive motion $r_{\perp}$ depends 
on $t$ and therefore the dephasing factor 
is of the type $\exp(-(t/\tau_{\varphi})^{(2{-}\sigma)/2})$.
In all other cases we have a simple exponential 
dephasing factor.

Obviously~(\ref{e34}) is a special case of the 
general formula~(\ref{e37}).  Note that~(\ref{e37})
should be multiplied by $\hbar^2 t$ in order to 
become a proper expression for $S_N$. 
By comparing~(\ref{e37}) to~(\ref{e34}) one 
can reveal the actual condition for the validity of the WNA. 
Within the interval $0 < k < 1/\ell$, the power 
spectrum of the motion occupies 
a frequency range $\omega<\omega_{cl}$, 
where the effective cutoff is 
\begin{eqnarray} \label{e38}  
\omega_{cl} \ = \ \left\{ \matrix{ 
v/\ell   & \mbox{if $k{=}1/\ell$ is located within the ballistic regime} \cr  
D/\ell^2 & \mbox{if $k{=}1/\ell$ is located within the diffusive regime}
} \right. 
\end{eqnarray}
As long as $\omega_{cl}<k_BT/\hbar$ one can 
use the WNA in estimating the integral in~(\ref{e37}). 
One observes that under such circumstances 
(\ref{e37}) reduces, after multiplication 
by $\hbar^2 t$, to~(\ref{e34}).

\section{Dephasing at Zero Temperatures}

As the temperature becomes low, such that 
$k_BT / \hbar < \omega_{cl}$, the dephasing 
rate becomes larger than the value which is 
predicted by the WNA.  
This is due to the "zero point fluctuations" (ZPF) in 
the frequency zone $k_BT/\hbar<\omega$.  
The temperature becomes effectively zero 
if $k_B T /\hbar < 1/t$. Time longer than $t$ 
is required in order to resolve such 
low temperatures. One can use then 
the zero temperature limit of $\phi(\tau)$ 
for which $\phi(\omega){=}\hbar\eta\omega$.    
This power spectrum corresponds to the ZPF 
of the environmental modes.

For either ballistic or chaotic motion,
The integral~(\ref{e37}) is dominated 
by $(k,\omega)$ modes that are concentrated 
below the curve $\omega=vk$. One obtains
\begin{eqnarray} \label{e39}  
\frac{1}{\tau_{\varphi}} \ \approx \  
\frac{C}{(1{+}\sigma)\pi}
\times\frac{1}{\hbar} \eta\ell \ v  
\ \ \ \ \ \mbox{for ballistic or chaotic motion}
\end{eqnarray}
Similarly, for diffusive motion the integral 
is dominated by $(k,\omega)$ modes that are 
concentrated below the curve $\omega=Dk^2$.
One obtains
\begin{eqnarray} \label{e40}  
\frac{1}{\tau_{\varphi}} \ = \   
\frac{C}{(2{+}\sigma)\pi}
\ln\left(1{+}\left(\frac{\ell v}{D}\right)^4\right)
\times\frac{1}{\hbar} \eta D \ 
\ \ \ \ \ \mbox{for diffusive motion}
\end{eqnarray}
In both cases most of the contribution comes 
from modes with large wavenumber, 
namely $k {\sim} 1/\ell$. 

The validity of the present semiclassical approach, 
which is based on the stationary-phase approximation 
is limited.  Common wisdom~\cite{wisdom} is that 
the applicability of the semiclassical approach is 
restricted to circumstances such that the energy 
transfer between the particle and the environment 
is much smaller than the particle's available energy. 
Technically, it is equivalent to the assumption of 
uncoupled wave-equations~\cite{uzy}. 
The {\em coupled} wave-equations for the 
particle-oscillators system can be {\em uncoupled} 
provided certain conditions are satisfied. 
Then we can treat the particle as moving with 
constant velocity $v$ and solve for the 
oscillators. It turns out that this reduction 
requires the assumption of small energy transfer.  
Therefore, one should anticipate problems once oscillators 
with $\omega_{\alpha}$ larger than $E$ are involved. 
In the latter case, there is no justification to think 
of the particle as decoupled from the bath, moving 
with some constant velocity, capable of exciting 
oscillators along the way. Therefore the 
corresponding factor $\exp(-S_N/\hbar^2)$ loses     
its physical significance. 
From the above considerations it follows that 
a reasonable condition for the validity of 
our semiclassical approach is 
\begin{eqnarray}    \label{e41}
\hbar\omega_{cl} < E 
\ \ \ \ \ \mbox{definition of large energy} 
\end{eqnarray}
On physical grounds this is the condition 
for being able to leave a trace "along the way".
The semiclassical significance of this condition can 
be further illuminated. For ballistic-like motion it is 
equivalent to the condition of small momentum 
transfer $\hbar/\ell < p$. This is precisely the 
condition for the applicability of the 
semiclassical methods for the scattering of the 
particle by the oscillators. More generally, 
one can define a quantum mechanical version 
of $P(k,\omega)$ as the FT of the 
correlator of the Heisenberg-picture operator    
$\exp(ik\bbox{x}(t))$. The quantal $P(k,\omega)$
coincides with the classical $P(k,\omega)$
only for $\omega<E/\hbar$. Larger classical 
frequencies are not supported by the effectively  
banded energy spectrum.      
Our practical conclusion is that the contribution 
of ZPF should be included if non-thermal motion with 
large energy is concerned. It should be excluded in case 
of thermal motion with $E \sim k_B T$.

\section{Dephasing due to Thermal Noise}

The possible contribution of ZPF to the dephasing 
has been discussed in the previous section. Now 
we are interested in the the thermal noise 
contribution (TNC). The TNC can 
be calculated using~(\ref{e37}) with $\phi(\omega)$
replaced by $\phi_T(\omega)$. See (\ref{e42}) for 
the definition of the latter. As the temperatures
is being raised various regimes are encountered:
\begin{eqnarray}  \nonumber
\mbox{Extremely Low Temperatures}  
& \ \ \ \ \ \Leftrightarrow \ \ \ \ \  
{1}/{t} & < {k_BT}/{\hbar} < {1}/{\tau_{\perp}}  
\\ \nonumber
\mbox{Low Temperatures}
& \ \ \ \ \ \Leftrightarrow \ \ \ \ \  
{1}/{\tau_{\perp}} & < {k_BT}/{\hbar} < \omega_{cl}  
\\ 
\mbox{High Temperatures}
& \ \ \ \ \ \Leftrightarrow \ \ \ \ \  
\omega_{cl} & < {k_BT}/{\hbar}  
\label{e43}
\end{eqnarray}
The time scale $\tau_{\perp}$ corresponds 
to the transverse distance $r_{\perp}$ of~(\ref{e35}).  
It equals $\bbox{d}/v$ for ballistic motion, 
and $\tau_{\tbox{erg}}$ for ergodic motion.
In case of non-ergodic diffusive motion, 
where $\tau_{\perp}=t$,  
the extremely low temperature regime is absent.
At high temperatures we can use the WNA as discussed 
in previous sections. The low and the extremely low 
temperature regimes are discussed in the next    
two paragraphs. 

When considering {\em extremely low} temperatures 
it is useful to define the critical 
exponent $\sigma_c=1$ for ballistic motion, 
and $\sigma_c=2$ for diffusive motion. 
If $\sigma<\sigma_c$ the integration in~(\ref{e37})
is dominated by $k\sim k_{\perp}$ and one obtains the result 
\begin{eqnarray} \label{e45}  
\frac{1}{\tau_{\varphi}} \ = \ 
{C'}\frac{2\eta k_BT }{\hbar^2} \ell^2 \times 
\left(\frac{\ell}{r_{\perp}}\right)^{\sigma} \times
\left(\frac{k_BT/\hbar}{1/\tau_{\perp}}\right)
\end{eqnarray}
For $\sigma_c<\sigma$ the integration is dominated 
by $k\sim 1/\ell$ and one obtains the result  
\begin{eqnarray} \label{e46}  
\frac{1}{\tau_{\varphi}} \ = \ 
{C'}\frac{2\eta k_BT }{\hbar^2} \ell^2 \times 
\left(\frac{k_BT/\hbar}{\omega_{cl}}\right)
\end{eqnarray}
In the formulas above $C'$ is a numerical 
factor of order unity. Note that in both 
cases, disregarding possible contribution of ZPF, 
the dephasing rate is proportional to~$T^2$. 
This should be contrasted with the high temperature
behavior where, disregarding the special case of 
non-ergodic diffusive motion, the dephasing rate 
is proportional to~$T$.

At {\em low} (but not extremely low) temperatures
it is useful to define a temperature 
dependent wavenumber as follows:
\begin{eqnarray}    \label{e44}
k_T \ = \ \left\{ \matrix{
(k_BT /\hbar v) \ \ \ & \mbox{in the ballistic regime} \cr
(k_BT/\hbar D)^{1/2}  & \mbox{in the diffusive regime}
} \right. 
\end{eqnarray}
For $\sigma<0$ the integration in~(\ref{e37}) 
is dominated by $k\sim k_{\perp}$,  
and one can use the WNA result~(\ref{e36}).
Thus, for $\sigma<0$, in the absence of ZPF contribution, 
we can trust the WNA at both high and low temperatures, 
and an actual crossover is expected only 
when extremely low temperatures are involved. 
For $0<\sigma<\sigma_c$ the integration is dominated 
by $k\sim k_T$, and one obtains the result  
\begin{eqnarray} \label{e47}  
\frac{1}{\tau_{\varphi}} \ = \ 
{C'}\frac{2\eta k_BT }{\hbar^2} \ell^2 \times 
(\ell \ k_T)^{\sigma}   
\end{eqnarray}
If $\sigma_c<\sigma$ 
the integration is dominated by $k\sim 1/\ell$ and 
one obtains again~(\ref{e46}).   
Thus, for $0<\sigma<\sigma_c$, there is a non-trivial 
low-temperature regime where the TNC to the dephasing rate 
is proportional to a non-universal power of $T$.   
Below this intermediate temperature-regime the TNC   
is proportional to~$T^2$. Above this intermediate 
temperature-regime we can trust the WNA and the 
dephasing rate is proportional to~$T$.

\section{Manifestation of Effective Static Disorder}

We shall consider in this section quantal Brownian 
motion in $2<d$ dimensions, that is described by the 
the DLD model with short range spatial correlations. 
For concreteness let us assume that the fluctuations 
of the effective stochastic potential satisfy:
\begin{eqnarray} 
\langle{\cal U}(\bbox{x}'',t''){\cal U}(\bbox{x}',t')\rangle
\ = \ \phi(t''{-}t')\cdot \ell^2\exp\left(-\frac{1}{2}
\left(\frac{\bbox{x}''{-}\bbox{x}'}{\ell}\right)^2\right)   
\end{eqnarray}
If we are interested in the dynamics over a 
{\em finite} time interval $t$, then obviously 
all the Fourier components in the frequency 
regime $|\omega|<1/t$ will have the same effect 
as static disorder. Let us denote by $\bar{{\cal U}}(\bbox{x},t)$ 
the "static" component of the effective stochastic 
potential. The variance of $\bar{{\cal U}}$ is 
determined by the product $\phi(\omega{=}0)\cdot (1/t)$. 
Recall that $\phi(\omega{=}0)\equiv\nu$. Consequently we have 
\begin{eqnarray} 
\langle\bar{{\cal U}}(\bbox{x}'',t'')
\bar{{\cal U}}(\bbox{x}',t')\rangle
\ = \ W^2 \ \exp\left(-\frac{1}{2}
\left(\frac{\bbox{x}''{-}\bbox{x}'}{\ell}\right)^2\right)   
\end{eqnarray}
where $W^2=\nu\ell^2/t$. We would like to find out 
whether this effective static disorder will 
manifest itself. Similar question has been 
raised in~\cite{felix}. One should not take for granted 
that the effect of low-frequency fluctuations is 
completely masked by the the incoherent effect 
of the high frequency modes.  

The first obvious step is to calculate the 
"statistical" dephasing rate due to the 
presence of the (effective) disorder. We can use 
(\ref{e37}) with the formal substitution 
$\phi(\tau)=W^2/\ell^2$, thus obtaining
\begin{eqnarray} 
\frac{1}{\tau_{\varphi}^0} \ = \ 
\frac{1}{\hbar^2}\frac{W^2}{\ell^2}
\int_0^{\infty} g(k)dk \ P(k,\omega{=}0)  
\end{eqnarray}
For ballistic-like motion one obtains 
\begin{eqnarray} 
\frac{1}{\tau_{\varphi}^0} \ = \ 
\frac{1}{\hbar^2} W^2 \frac{\ell}{v} \ = \
\left(\frac{\nu\ell^3}{\hbar^2 v}\right) 
\cdot \frac{1}{t}
\end{eqnarray}
In order to have a non-vanishing effect we 
should have $\tau_{\varphi}^0 \ll t$. This 
condition can be cast into the form 
$\xi \ll vt$ where $\xi=(\hbar v)^2/(\ell W^2)$ 
is the mean free path. Still another way 
to express this condition is  
$\tau_{\varphi}^{\tbox{WNA}} \ll \ell/v$, 
where $\tau_{\varphi}^{\tbox{WNA}}$ is given by~(\ref{e28}).   
On the over hand, the actual dephasing 
time should satisfy $\ell/v < \tau_{\varphi}$, 
else coherent effects due to the scattering 
by the (effective) disordered potential  
will not manifest themselves. Thus, we come to 
the conclusion that the following condition 
should be satisfied in order to have manifestation 
of coherent effects due to scattering by the 
effective static disorder: 
\begin{eqnarray} 
\tau_{\varphi}^{\tbox{WNA}} 
\ll \frac{\ell}{v} 
\ll \tau_{\varphi}
\end{eqnarray}
Obviously, this condition can be satisfied only 
at the low temperature regime.

\section{Dephasing versus Dissipation, Concluding Remarks}

It is quite striking that the friction 
coefficient $\eta$ is not affected by 
quantum mechanical effects. 
Having the Boltzmann picture in mind 
it is not anticipated that all the 
quantum mechanical scattering events 
will conspire to give the classical 
result. Still, this conclusion follows 
from the FV-formalism quite easily.      
In order to have a clear physical picture,  
let us consider the time evolution of 
a Gaussian wavepacket using either 
the propagator~(\ref{e24}) or the 
equivalent Master equation~(\ref{e25}). 
Clearly, the dissipation rate $dE/dt$ 
is the same as in the classical picture, with 
small transient corrections. 
Thus, also in the quantum-mechanical picture, 
the damping process is characterized by 
the time constant $\tau_{\eta}=(\eta/m)^{-1}$. 

Consider a ballistic-like motion with 
non-thermal energy, such that $\hbar/\ell \ll p$.
Consequently the equivalent condition~(\ref{e41}) is 
satisfied, and we can trust semiclassical considerations   
as far as dephasing is concerned.   
We can cast the universal WNA result~(\ref{e28}) and 
the ZPF result~(\ref{e39}) into the form  
\begin{eqnarray} 
\frac{1/\tau_{\varphi}}{1/\tau_{\eta}} \ \approx \
\left\{ \matrix{ 
\left({\ell}/{\lambda_T}\right)^2 & \mbox{high-T} & \mbox{(universal)} \cr
\left({\ell}/{\lambda_B}\right)   & \mbox{low-T}  & \ 
}\right.
\end{eqnarray}
where $\lambda_T$ is the thermal wavelength, 
and $\lambda_B$ is the De-Broglie wavelength. 
We see that for the above {\em non-thermal motion}      
the dephasing time is always shorter than the 
damping time.  The dephasing rate is linear in $T$ at 
high temperature, and saturates at low temperatures.   

For {\em thermal motion} the latter statement is no-longer 
true. At the low temperature regime the ZPF contribution 
should be excluded, and therefore the dephasing rate
goes to zero, while the damping time remains finite. 
One should distinguish between various physical 
circumstances. At extremely low temperatures, assuming 
ergodic motion, the dephasing rate is proportional to 
$T^2$. At higher temperatures we can trust the WNA 
provided $\sigma<0$. Otherwise, there is an intermediate 
low-temperature regime where the dephasing rate is  
proportional to a non-trivial power of $T$. 

In this paper we have introduced a systematic derivation 
of a general formula for the dephasing rate, Eq.(\ref{e37}), 
that holds in all physical circumstances, including the ZCL limit.
Expressions similar to Eq.(\ref{e37}), that incorporate 
integration over the $(k,\omega)$ environmental 
modes, are encountered frequently in the literature, 
starting from the well known work in~\cite{schmid}. 
Whenever diffusive motion of electrons is concerned
(see Appendix E), our expressions agree with well know 
results~\cite{imry}.     
However, most publications avoid a straightforward application 
of the FV formalism, and introduce in some stage heuristic 
considerations in order to obtain a convergent result. 
There is a "zoo" of cutoffs that are introduced 
in performing the $(k,\omega)$ integration, some of them 
are questionable. For example, it is customary to take 
$1/\tau_{\varphi}$ itself as a lower cutoff for the 
$\omega$ integration. The upper cutoff is sometimes 
$k_BT/\hbar$, sometimes Fermi energy, sometimes 
the kinetic energy that corresponds to the drift velocity, 
sometimes the inelastic scattering rate, and so on. 
The role of Fermi statistics in the determination of 
the various cutoffs is usually left unclear. There are 
similar ambiguities in the determination of the the proper 
cutoffs for the $k$ integration.  Our derivation 
has led to a proper definition for the power-spectrum of 
the motion, and has made it unnecessary to introduce ad-hoc 
cutoffs into the calculations.  Furthermore, in the 
analysis of dephasing at the limit of zero temperature,    
we were successful in reducing our considerations to the 
level of "one particle physics", thus avoiding a complicated 
discussion of the role played by Pauli exclusion principle.


\ack{
I thank Uzy Smilansky and Yoseph Imry for useful 
discussions. Some of the motivation for this 
work has come following suggestions by Yuval Gefen.  
This research was supported by the Minerva Center 
for Nonlinear Physics of Complex systems.
}



\appendix

\section{Useful Identity}

We first cite the well known identity
\begin{eqnarray} \nonumber
\hspace*{-1cm}
\frac{1}{\sqrt{2\pi\delta t}} 
\exp\left(+\frac{i}{2\delta t}(x-x_0)^2\right) 
\ = \ 
\left[1+\frac{i}{2}\delta t \partial_x^2 
+{\cal O}(\delta t^2)\right] \delta(x-x_0) 
\end{eqnarray}
Both sides of this identity should be 
interpreted as kernels of operators. 
When applied to wave-functions, the left hand 
side corresponds to the free-motion propagator, 
and the second term in the right hand side 
corresponds to the free-motion Hamiltonian.  
We shall explain now how to derive 
the following related identity:
\begin{eqnarray} \nonumber
\frac{1}{2\pi\delta t} 
\exp\left(\frac{i}{\delta t}(R-R_0)(r-r_0)
-i\eta{\cdot}(R-R_0) \right) 
\ = \  \\ 
\left[1+i\delta t \partial_R\partial_r
+\eta \delta t \partial_r 
+ {\cal O}(\delta t^2) \right] 
\delta(R-R_0)\delta(r-r_0) 
\end{eqnarray}
Both sides of the latter identity should be 
interpreted as kernels of operators that 
operate on phase-space functions.
We start the derivation by writing an equality  
that follows from the first identity via
simple replacements:  
\begin{eqnarray} \nonumber
\hspace*{-1cm}
\frac{1}{\sqrt{2\pi\delta t}} 
\exp\left(\pm\frac{i}{2\delta t}
(x_{\pm}\mp \half\eta\delta t))^2\right) 
\ = \ 
\left[1+\frac{i}{2}\delta t \partial_{\pm}^2 
+{\cal O}(\delta t^2)\right] 
\delta(x_{\pm}\mp\half\eta\delta t) 
\end{eqnarray}
Upon multiplication it follows that 
\begin{eqnarray} \nonumber
\hspace*{-1cm}
\frac{1}{\sqrt{2\pi\delta t}} 
\exp\left(\pm\frac{i}{2\delta t}
(x_{+}^2-x_{-}^2)- 
i\frac{1}{2}\eta(x_{+}+x_{-})\right) 
\ = \\ \nonumber 
\hspace*{-1cm}
\left[1+\frac{i}{2}\delta t \partial_{+}^2\right]
\left(\delta(x_{+})-\frac{1}{2}\eta\delta t\delta'(x_{+})\right) 
\cdot
\left[1-\frac{i}{2}\delta t \partial_{-}^2\right]
\left(\delta(x_{-})+\frac{1}{2}\eta\delta t\delta'(x_{+})\right)
\end{eqnarray}
Simplification of the right hand side gives
\begin{eqnarray} \nonumber
\hspace*{-1cm} 
= \ \left[1+\frac{i}{2}\delta t (\partial_{+}^2-\partial_{-}^2)
+\frac{1}{2}\eta\delta t (\partial_{+}-\partial_{-}) 
+{\cal O}(\delta t^2)\right]
\delta(x_{+}) \delta(x_{-})  
\end{eqnarray}
The replacements $x_{+}\rightarrow(x''-x_0'')$ and
$x_{-}\rightarrow(x'-x_0')$, followed by 
the transformation to the variables 
$R=(x''+x')/2$ and $r=(x''-x')$, gives the 
desired result. Note that the derivation holds 
also in the case where $\eta$ is replaced 
by some function of $r$.

\section{Derivation of the Master Equation}

Wigner function $\rho(R,P)$ is the 
Fourier transform of the reduced probability 
density matrix $\rho(R,r)$ in the variable 
$r\leadsto P$. The path integral expression 
for the kernel ${\cal K}(R,r|R_0,r_0)$ is the  
type $\int\int{\cal D}R{\cal D}r$ with
an obvious endpoint conditions, and 
its integrand is essentially the same as in 
(\ref{e18}). See~\cite{dld} for more details. 
The infinitesimal-time kernel equals 
simply to the integrand of the path 
integral expression, namely
\begin{eqnarray} \nonumber
\hspace*{-2cm} 
{\cal K}_{\delta t} \ = \ \exp\left( 
\frac{i}{\delta t}\frac{m}{\hbar}(R-R_0)(r-r_0)
+i\frac{\eta}{m}w'(r){\cdot}(R-R_0) 
-\delta t \frac{\nu}{\hbar^2}(w(0)-w(r)) 
\right) \ = \ \\ \nonumber 
\hspace*{-2cm} 
\left[ 1+ \delta t \left(
i\frac{\hbar}{m}\partial_R\partial_r 
-\frac{\eta}{m}w'(r)\partial_r 
-\frac{\nu}{\hbar^2}(w(0)-w(r)) 
\right) + {\cal O}(\delta t^2) \right]
\delta(R-R_0)\delta(r-r_0)   
\end{eqnarray}
where the second line is obtained by employing 
the identity of Appendix A. Thus, to 
leading order, the kernel 
${\cal K}_{\delta t}$  has the same 
effect as operating with a differential  
operator of the type $1+\delta t{\cal L}$. 
Consequently, we find that $\rho(R,r)$ satisfies 
a Master equation of the form 
${\partial\rho}/{\partial t} = {\cal L}\rho$. 
This Master equation is easily transformed 
into the Wigner representation, where
any derivative $\partial_r$ is replaced by 
multiplication with $iP/\hbar$. Similarly,     
any multiplication by a (real symmetric) function 
$\tilde{G}(r)$ is transformed into a convolution 
with a (real symmetric) kernel $G(P{-}P')$. 
For the convolution $\int G(P{-}P') \rho(P') dP'$ 
we use the notation $G\star\rho$. 
A kernel $G(P{-}P')$ and a function 
$\tilde{G}(r)$ are related by a Fourier 
transform with the convention
\begin{eqnarray} \nonumber
G(P{-}P') = \frac{1}{2\pi\hbar} \int_{-\infty}^{\infty}
G(r) \ \cos\left( \frac{P{-}P'}{\hbar}r \right) \ dr
\end{eqnarray}
In the above expression we have modified the standard 
FT convention, by including the factor $1/(2\pi\hbar)$. 
This has been done in order to have properly normalized 
kernels, with the measure $dP$ rather than $dP/(2\pi\hbar)$.  
Note also that the FT of $w'(r)$ equals 
to the $i\hbar\partial_P$ derivative of 
the kernel $G_F$ that corresponds to 
the real symmetric function $w'(r)/r$.

\section{Dephasing and Inelastic Scattering}

Consider the interference contribution of two 
trajectories $a\&b$ that are related by time reversal.
Assuming short range interaction with the 
environmental modes we can prove that dephasing 
is related to the probability for leaving a trace
in the environment. This statement is true at any 
temperature. "Short range interaction" means that 
$0<\sigma$ and that $\ell$ is a small scale.
"Leaving a trace" means that at least one of the 
oscillators has changed its quantum-mechanical state.  
It follows that under such circumstances 
dephasing-rate is equal to the inelastic scattering 
rate.  

Recall the definition of the influence functional.
For simplicity we consider first a zero-temperature 
bath, meaning that all the bath-oscillators are initially 
in the ground state. These oscillators are driven by the 
motion of the particle. The excited states of the 
bath will be denoted by $| \{n_{\alpha}\} \rangle$. 
The evolution operator of a driven oscillator 
will be denoted by $U_{\alpha}[x]$.   
\begin{eqnarray} \nonumber
\hspace*{-1cm}
F[\bbox{x}_a,\bbox{x}_b] \ \equiv \ 
\sum_{ \{n_{\alpha}\} } \ 
\left\langle \{n_{\alpha}\} \left| 
\prod_{\alpha} U_{\alpha}[\bbox{x}_b] 
\right| \{0\} \right\rangle
\left\langle \{n_{\alpha}\} \left| 
\prod_{\alpha} U_{\alpha}[\bbox{x}_a] \right| 
\{0\} \right\rangle^{\star} 
\end{eqnarray}
If $a\&b$ are related by time reversal, then one can consider 
only those oscillators that are located "along the loop". Each 
oscillator along the loop is characterized by its 
natural frequency $\omega_{\alpha}$ and by the time 
$t_{\alpha}$ at which $x_a(t) \approx x_{\alpha}$. 
Consequently we have the following expression for the 
influence functional:
\begin{eqnarray} \label{e201}
F[\bbox{x}_a,\bbox{x}_b] \ = \ 
\sum_{\{n_\alpha\}} \ \mbox{e}
^{-i\sum_{\alpha}n_{\alpha}\omega_{\alpha}\cdot(2t_{\alpha}{-}t)}
\ \ P(\{n_\alpha\}|\{0\})
\end{eqnarray}
where the excitation probability is 
$P(\{n_\alpha\}|\{0\})\equiv\prod
|\langle n_{\alpha}| U_{\alpha}[\bbox{x}_a] | 0 \rangle|^2$. 
We would like to argue that 
$F[\bbox{x}_a,\bbox{x}_b]=P(\{0\}|\{0\})$.  
Indeed, the summation in (\ref{e201}) contains  
\mbox{\em one-oscillator} excitations for which $\sum n_{\alpha}=1$, 
\mbox{\em two-oscillator} excitations for which $\sum n_{\alpha}=2$ 
and so on. Let us consider all the 
\mbox{one-oscillator} excitations that involves  
$\omega_{\alpha}$ in the range $[\omega,\omega{+}\delta\omega]$. 
Each of these excitations contributes 
the same $P(\{n_{\alpha}\}|\{0\})$, but with 
a different phase factor $\exp(i2\omega_{\alpha}t_{\alpha})$. 
By construction of the DLD model (Eq.(\ref{e11})), 
the summation over the phase factors will lead to 
a zero contribution. The meaning of "zero contribution"    
is as follows: either we average over realizations of 
$x_{\alpha}$, or else we recall that the influence functional 
appears inside a path-integral expression. The summation 
over paths will have the same effect like an averaging procedure, 
and therefore we will have indeed a "zero contribution" in the 
\mbox{mathematically-generalized} sense.  Similar argumentation 
applies to other subsets of excitations.   
If the bath is initially in thermal equilibrium, rather than 
in zero temperature, than the relation  
$F[\bbox{x}_a,\bbox{x}_b]=P(\{0\}|\{0\})$ can be generalized 
into 
\begin{eqnarray} \label{e202}
F[\bbox{x}_a,\bbox{x}_b] \ = \ 
\exp(-S_N[\bbox{x}_a,\bbox{x}_b]/\hbar^2) 
\ = \ P(\mbox{leaving no trace}) \ \ \ .
\end{eqnarray}
In order to prove the latter statement one should 
use the same procedure as above with $|\{0\}\rangle$ replaced 
by some arbitrary initial preparation $|\{n^0_\alpha\}\rangle$. 
Then the result should be thermally averaged.

\section{Statistical Characterization of the Trajectories}

Having a pair $a\&b$ of trajectories, we can
define the function   
\begin{eqnarray} \label{e100}
P_{ab}(\bbox{r},\tau)=\langle\delta(\bbox{r}-
(\bbox{x}_a(t'{+}\tau)-\bbox{x}_b(t' )))\rangle
\end{eqnarray}
where the average is over $t'$ within the time interval 
$[0,t]$ which is considered. We shall use the 
notations $P_{aa}=P_{\parl}$ and 
$P_{ab}=P_{\perp}$ for $a\ne b$.
It is also useful to define the functions
\begin{eqnarray} \nonumber  
P(\bbox{r},\tau) \ \equiv \ 
P_{\parl}(\bbox{r},\tau)-P_{\perp}(\bbox{r},\tau) 
\\ 
P(\bbox{r}) \ \equiv \ P(\bbox{r},0).
\label{e101}
\end{eqnarray}
The Fourier transform of $P(\bbox{r})$ will be denoted 
by $P(\bbox{k})$, and the double Fourier transform of 
$P(\bbox{r},\tau)$ will be denoted by $P(\bbox{k},\omega)$. 
The above functions all appear within integrals 
where "isotropic" integration over $\bbox{r}$ or $\bbox{k}$
is being performed. Therefore it is convenient 
to average all these functions over all orientations,  
and so to have functions that depend on 
either $r=|\bbox{r}|$ or $k=|\bbox{k}|$ respectively. 
A useful notation is 
$\mbox{Cos}(\bbox{r})\equiv\langle
\cos(\bbox{\Omega}{\cdot}\bbox{r})\rangle$, 
where the average is over the orientation 
of the unit vector $\bbox{\Omega}$. The 
function $\mbox{Cos}(\bbox{r})$ depends only 
on $|\bbox{r}|$. It equals $\cos(r)$ in 1-D, 
regular Bessel function ${\cal J}_0(r)$ 
in 2-D, and $\mbox{sinc}(r)$ in 3-D.

We shall distinguish now between {\em ballistic} 
trajectories as in Fig.\ref{f_trajc}(a), 
{\em chaotic} trajectories as in Fig.\ref{f_trajc}(b),
and {\em diffusive} trajectories as in Fig.\ref{f_trajc}(c).
We can also add to this list the case 
of diffusive trajectories that cover ergodically 
the whole available space.  
In case of a ballistic trajectory 
$P_{\parl}(\bbox{r},\tau)=\bbox{\delta}(\bbox{r}-\bar{\bbox{r}})$, 
leading to $P_{\parl}(\bbox{k},\tau)=\mbox{Cos}(k\bar{r})$, 
where $\bar{r}=v\tau$. Similar expression hold 
for $P_{\perp}$, where $\bar{r}\approx(\bbox{d}^2+(v\tau)^2)^{-1/2}$, 
and $\bbox{d}$ is the transverse distance between the slits. 
For a chaotic trajectory $P_{\parl}(r,\tau)$
starts as in the ballistic case,
while $P_{\perp}(r,\tau)$ is uniform in $\bbox{r}$.
The ergodic time is given essentially by the 
ballistic time, namely $\tau_{\tbox{erg}}=\tau_{\tbox{bal}}=L/v$, 
where $L$ is the linear dimension of the cavity. 
For $\tau_{\tbox{erg}}<\tau$ both $P_{\parl}(r,\tau)$ and
$P_{\perp}(r,\tau)$ become uniform in $\bbox{r}$. 
For diffusive motion $P_{\parl}(\bbox{r},\tau)$ 
is as for the ballistic case as long as 
$\tau<\tau_{\tbox{bal}}$ where $\tau_{\tbox{bal}}=D/v^2$. 
On larger times it becomes a Gaussian, and consequently 
$P_{\parl}(\bbox{k},\tau)=\exp(-Dk^2\tau)$,   
where $D$ is the diffusion coefficient. 
The latter expression as well as the approximation 
$P_{\perp}(\bbox{r},\tau) \approx P_{\parl}(\bbox{r},t)$  
hold for non-ergodic diffusive motion. After 
the ergodic time $\tau_{\tbox{erg}}=L^2/D$, both 
$P_{\parl}(r,\tau)$ and $P_{\perp}(r,\tau)$ become 
uniform in $\bbox{r}$.  

In the following paragraphs we discuss the 
various $k$ regimes of the function $P(k,\omega)$. 
We shall distinguish between the small-$k$ 
regime and the large-$k$ regime. In case 
of diffusive trajectories the large-$k$ regime 
will be further divided into a ballistic regime 
($v/D<k$) and a diffusive regime ($k<v/D$), where 
$v/D$ is the inverse of the mean free path.    
We turn first to define the notions of small 
and large $k$. The transverse distance between 
two trajectories has a distribution $P_{\perp}(r)$, 
and we can define a typical value 
$r_{\perp}=(\langle \bbox{r}^2 \rangle)^{1/2}$. 
See Eq.(\ref{e35}). The associated wavenumber 
is $k_{\perp} \equiv 1/r_{\perp}$.  
Large $k$ means from now on $k_{\perp} \ll k$.  
For large $k$ we may use the approximation
\begin{eqnarray} \label{e102}
P(k,\omega) \ \approx \ P_{\parl}(k,\omega)
\ \ \ \ \ \mbox{for large $k$ \ (meaning $k_{\perp} \ll k$)}
\end{eqnarray}
For either ballistic, diffusive or   
chaotic trajectories we have 
\begin{eqnarray} \label{e103}  
P(k,\omega) \approx \frac{1}{vk} 
\hat{B}\left(\frac{\omega}{vk}\right)
\ \ \ \ \ \mbox{in the ballistic regime} 
\end{eqnarray}   
where $\hat{B}$ is the FT of the $Cos$ function. 
This rectangular-like scaling-function has a 
unit width and a unit normalization. 
In case of diffusive motion we also have 
\begin{eqnarray} \label{e104}  
P(k,\omega) \ \approx \ \frac{2Dk^2}{(Dk^2)^2+\omega^2}
\ \ \ \ \ \mbox{in the diffusive regime} 
\end{eqnarray}   
Diffusive regime means here any large $k$ 
that satisfies $k<v/D$. The validity of the 
latter approximation is further restricted by 
the condition $\omega<v^2/D$. The collision frequency 
$\omega \sim v^2/D$ should be used as a cutoff to the 
slow $1/\omega^2$ power-law decay.   
In our calculations we shall assume 
that $1/\ell \ll v/D$, meaning that 
the motion is diffusive also on the 
spatial scale $\ell$.

For small $k$, meaning $k \le k_{\perp}$,  
the transverse term cannot be ignored, 
and we no-longer can use the approximation 
(\ref{e102}). We can define a time scale 
$\tau_{\perp}$ that correspond to $k_{\perp}$. 
It equals $\bbox{d}/v$ for ballistic motion, 
$t$ for diffusive motion, and $\tau_{\tbox{erg}}$ for 
ergodic motion. For a given small $k$, the 
function $P(k,\tau)$ is concentrated 
within $\tau<\tau_{\perp}$, since $P_{\parl}$ 
and $P_{\perp}$ are not identical there. 
Consequently      
\begin{eqnarray} \label{e105}
P(k,\omega) \ \approx \ \tau_{\perp} {\cdot} P(k) \ \ \ 
\mbox{for $\omega < 1/\tau_{\perp}$ in the small $k$ regime}
\end{eqnarray}
The function $P_{\perp}(\bbox{k})$ is the FT of 
$P_{\perp}(\bbox{r})$, and therefore we have   
for $P(k)=1{-}P_{\perp}(\bbox{k})$ the following 
small $k$ approximation 
\begin{eqnarray} \label{e106}
P(k) \ \approx \  
\frac{1}{2} r_{\perp}^2 k^2
\ = \ \left\{\matrix{ 
\half \bbox{d}^2 k^2 & \mbox{for ballistic motion} \cr
Dt{\cdot}k^2 & \mbox{for diffusive motion} \cr
\half L^2 k^2 & \mbox{for ergodic motion} 
}\right.
\end{eqnarray}
For large $k$ we have $P(\bbox{k}) \approx 1$, 
as implied by~(\ref{e102}) and the definition~(\ref{e101}).

\section{Friction Constant For Electrons in Metal}

The effect of electron-electron Coulomb interaction 
can be analyzed by considering the motion of a single 
electron under the influence of a fluctuating  
electrostatic potential ${\cal U}(\bbox{x},t)$
that is created by the other electrons. 
Thus the electron experience an interaction with 
a fluctuating field that is characterized by  
$\langle{\cal U}(\bbox{x}'',t''){\cal U}(\bbox{x}',t')\rangle$.  
It is well known, using fluctuation-dissipation theorem, 
that for diffusive electrons the corresponding 
fluctuation spectrum is 
\begin{eqnarray} \nonumber 
\tilde{w}(\bbox{k}) \phi(\omega) \ = \ 
\frac{e^2}{\bbox{\sigma}_d}
{\cdot}\frac{1}{k^2}\omega 
\ \hbar\coth\left(\frac{\hbar\omega}{2k_BT}\right)
\ \ \ \ \mbox{for} 
\ \omega<\frac{1}{\tau_c}, 
\ \  |k|<\frac{1}{\ell}  
\end{eqnarray}   
The ohmic behavior is cut-off by the 
Drude collision frequency $1/\tau_c$. 
The elastic mean free path is $\ell=v\tau_c$, 
where $v$ is the Fermi velocity. 
For $k<1/\ell$ the power spectrum $P(k,\omega)$
of the diffusive motion is concentrated 
below the Drude cutoff frequency.   
The ballistic regime $1/\ell<k$ is of 
no interest because its contribution 
is suppressed due to the Drude cutoff.
One observes that these fluctuations 
corresponds to the generalized DLD 
model~(\ref{e14}) with $\sigma=d{-}2$.     
Having observed that the mean free path 
is the physical (effective) cutoff for the 
spatial fluctuations (this statement is true 
for $\sigma{<}4$), it follows 
from the convention $|w''(0)|=1$ that 
the friction constant (up to dimensionless 
factor of order unity) is given by  
the expression 
\begin{eqnarray} \label{e107}  
\eta \ = \  
\frac{e^2}{\bbox{\sigma}_d}
\left(\frac{1}{\ell}\right)^d 
\ = \ 
\frac{1}{D}\Delta_{\ell}
\end{eqnarray}   
Here $e$ is the charge of an electron, 
and $\bbox{\sigma}_d$ is the conductivity 
defined for $d=1,2,3$ dimensions. 
The last equality is obtained by using 
Einstein relation in order to express 
$\bbox{\sigma}_d$ in terms of $D$.
The notation $\Delta_{\ell}$ stand for 
the mean level spacing within a cube 
whose volume is $\ell^d$.


\ \\ \ \\

\noindent {\bf References} \\


\end{document}